# Reconfigurable on-chip entangled sources based on lithium-niobate waveguide circuits

H. Jin[1], F. M. Liu[2], P. Xu[1]*, J. L. Xia[2], M. L. Zhong[1], Y. Yuan[1], Y. X. Gong[3], W. Wang[2], S. N. Zhu[1]

[1]National Laboratory of Solid State Microstructures and College of Physics, Nanjing University, Nanjing 210093, China.

[2]Beijing Institute of Aerospace Control Devices, Beijing, 100094, China

[3]Department of Physics, Southeast University, Nanjing, 211189, China.

*Corresponding author, email: pingxu520@nju.edu.cn (P. Xu);

Integrated quantum optics becomes a consequent tendency towards practical quantum information processing. Here, we report the on-chip generation and manipulation of photonic entanglement based on reconfigurable lithium niobate waveguide circuits. By introducing periodically poled structure into the waveguide interferometer, two individual photon-pair sources with controllable phase-shift are produced and cascaded by a quantum interference, resulting in a deterministically separated identical photon pair. The state is characterized by 92.9±0.9% visibility Hong-Ou-Mandel interference. Continuous morphing from two-photon separated state to bunched state is further demonstrated by on-chip control of electro-optic phase-shift. The photon flux reaches ~$1.4\times10^7$ pairs $nm^{-1}$ $mW^{-1}$. Our work presents a scenario for on-chip engineering of different photon sources and paves a way to the fully integrated quantum technologies.

Numerous progresses have been achieved in integrated photonic circuits[1-10], providing a solid strategy for high-performance quantum information processing. However, most of the photonic chips require external photon sources, which inevitably hinder the integration complexity, therefore further integration of photon sources together with the photonic circuits are of essential importance[11-13]. A solid strategy for achieving this multifunctional integration turns to the periodically poled lithium niobate waveguide circuits.

The lithium niobate (LN) crystal, also called the 'silicon of photonics'[14,15], acts as one of the most versatile and widely used materials in integrated optical devices, owing to strong $\chi^{(2)}$ nonlinearity, large piezoelectric, acousto-optic and electro-optic coefficients features, as well as well-developed waveguide fabrication technique using either proton-exchange or titanium-indiffusion method. In telecommunications applications, the LN modulators based on electro-optic effect have reached 40 GHz (100 GHz in the laboratory[16]) and have recently been demonstrated for fast path and polarization control of single photons[17]. More specifically, domain-engineering technique[18,19] can be introduced to LN, leading to desirable quasi-phase-matching (QPM) spontaneous parametric downconversion (SPDC) process for efficient generation of entangled photons and possible extension of photon wavelength within the wide transparency window of LN. When complex designs of poling structure are considered, multiple operations on entangled sources can be achieved on a single bulk crystal platform[20,21]. The bulk LN can be further fabricated into waveguide structures, and the photon flux will be greatly enhanced as first demonstrated from a single periodically poled LN (PPLN) waveguide source[22,23]. All the aforementioned features allow complex waveguide circuits embedded with high-flux photon sources and reconfigurable phase shifters on a single LN chip, which makes LN qualified for more complex quantum tasks[6].

An identical photon pair, namely two photons sharing the same characters in all the degrees of freedom, plays as the key source in quantum communication and linear optical quantum computing. Here based on a single LN photonic chip we demonstrate the generation, deterministic separation and continuous path-entanglement evolution of such photon pair. The LN chip is sketched in Fig. 1. Basically it is composed by annealed proton exchanged channel waveguides integrated on a Z-cut periodically poled lithium niobate (PPLN) crystal. This design enables the generation, interference and filtering of entangled photons from separate regions of PPLN waveguides, leading to reconfigurable on-chip quantum light sources. The chip can be characterized by three sections. Section I is designed to deal with the classical pump light. A 780 nm pump is coupled into waveguide L$_0$ and equally distributed by a Y-branch beamsplitter, which is composed of single mode waveguides at the wavelength of 780 nm. After the Y-branch, electro-optic effect is considered to control the phase-shift between two paths. Electrodes are fabricated above two pump waveguides for applying a voltage. Transition tapers then follow to connect the 780 nm Y-branch with the 1560 nm single mode waveguides. Section II is the PPLN region, in which degenerate photon pairs at 1560 nm are generated indistinguishably from either one of the two PPLN waveguides, yielding a path-entangled state $\left(|2,0\rangle+e^{i\Delta\varphi}|0,2\rangle\right)/\sqrt{2}$, in which the relative phase $\Delta\varphi$ is transferred from the phase difference of pump modes. In Sect. III, quantum interference is realized by a $2\times2$ directional coupler ($C_1$). The on-chip phase is adjusted by changing the voltage applied in Sect. I. The entangled photons are separated from the pump by on-chip wavelength filters $C_2$ and $C_2'$. Entangled photons are transferred to the neighboring waveguides $R_1$ and $R_4$. The pump is left in $R_2$ and $R_3$. The whole chip is directly connected with optical fiber tips, which are fixed with the chip by using UV-curing adhesive after reaching a high coupling efficiency. The input waveguide $L_0$ is connected with a 780 nm single mode fiber (SMF) and output waveguides $R_1$ and $R_4$ are connected with 1560 nm SMFs. This removes the need for optimizing the coupling during the observation of quantum interference.

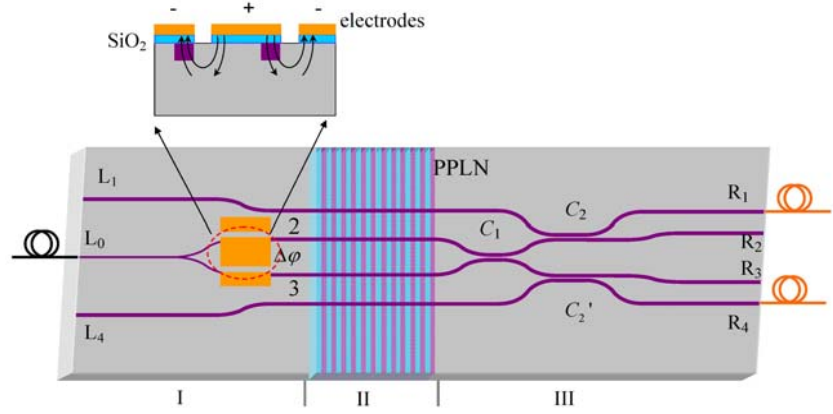

FIG. 1 Structure of the LiNbO$_3$ photonic chip. The photonic chip is structured on a Z-cut PPLN crystal of 0.5 mm thickness. The input pump is first equally split by a Y branch into two modes (in waveguides 2 and 3) with a phase difference $\Delta\varphi$, which is controlled by applying a bias voltage to the following electrodes (I). In the 10 mm long periodically poled section with a period of 15.32 μm, 1560 nm photon pairs are generated (II). Quantum interference is realized by the 2×2 directional coupler ($C_1$), and then the entangled photons are coupled to waveguides R$_1$ and R$_4$ by couplers $C_2$ and $C_2'$, leaving the pump in R$_2$ and R$_3$ (III). The input (L$_0$) and output (R$_1$ and R$_4$) waveguides are connected with fiber tips. The inset is the structure of electrodes, with the buffer layer (SiO$_2$) separated between the electrodes to suppress DC drift.

When the phase-shift $\Delta\varphi$ between pump modes in waveguide 2 and 3 is zero, the chip worked as a balanced time-reversed Hong-Ou-Mandel (BRHOM) interferometer, realizing a deterministic separation of an identical 1560 nm photon pair,

$$\frac{1}{\sqrt{2}}(|2,0\rangle+|0,2\rangle) \xrightarrow{BRHOM} |1,1\rangle. \tag{1}$$

In this case the photon pair always emit from different waveguides, one from $R_1$ and the other from $R_4$. The state is marked as $|\Psi\rangle_{separated}=|1,1\rangle$.

In a more general case, when the electro-optic phase-shift (EOPS) $\Delta\varphi$ is introduced, the state evolution in the chip can be written as

$$\frac{1}{\sqrt{2}}\left(|2,0\rangle+e^{i\Delta\varphi}|0,2\rangle\right)\xrightarrow{EOPS}\frac{1}{\sqrt{2}}(|2,0\rangle-|0,2\rangle)\sin(\Delta\varphi/2)+|1,1\rangle\cos(\Delta\varphi/2). \tag{2}$$

This can be addressed by quantum interference of two coherent photon-pair sources on the $2\times2$ coupler ($C_1$). The output state is a superposition of two-photon separated state $|\Psi\rangle_{separated} = |1,1\rangle$ ($\Delta\varphi = 0$) and bunched state $|\Psi\rangle_{bunched} = \frac{1}{\sqrt{2}}(|2,0\rangle - |0,2\rangle)$ ($\Delta\varphi=\pi$, the photon pair always emit together from either $R_1$ or $R_4$). By varying the bias voltage, thus the phase difference, we will observe a continuous evolution from two-photon separated state to bunched state. Specifically, the photon-pair rate from $R_1$ and $R_4$ follows

$$P_{separated} \propto \cos^2(\Delta\varphi/2) = \cos^2(\pi U/2V_\pi) \ , \qquad (3)$$

while the photon-pair rate from $R_1$ or $R_4$ gives

$$P_{bunched} \propto \sin^2(\Delta\varphi/2) = \sin^2(\pi U/2V_\pi) . \qquad (4)$$

The phase difference between two-path is calculated as $\Delta\varphi = 2\pi\Gamma\gamma_{33}n_p^3 LU/(\lambda_p d)$ when applying voltage $U$ on the pump waveguides. It is directly proportional to the voltage. The half-wave voltage is written as $V_\pi = \frac{\lambda_p d}{2\Gamma\gamma_{33}n_p^3 L}$, in which $\Gamma$ is a fill factor reflecting the overlap between the electric field and optical mode field[24], $\gamma_{33}$ is the electro-optic coefficient, $n_p$ is the refraction index of pump, $L$ and $d$ are the length and separation of electrode pairs. The period of interference fringes should be twice of the half-wave voltage i.e. $2V_\pi$.

In the following experiment, we first make a coincidence counting measurement on $R_1$ and $R_4$ to record how the bias voltage $U$ affects the output state. The experimental setup is sketched in Fig. 2a. The photonic chip is temperature controlled at 25.5 °C. The 780 nm pump is a continuous-wave fiber laser which is polarization controlled and then connected with input fiber $L_0$. Thanks to the 10 mm long periodically poled section with a poling period of 15.32 μm, degenerate photon pairs at 1560 nm are produced. We make coincidence counting measurement between $R_1$&$R_4$, $R_1$&$R_1$ and $R_4$&$R_4$. The coincidence counts are recorded in Fig.2(b), 2(c) and 2(d), respectively, exhibiting interference fringes with the visibility of 98.9±0.3%, 97.5±0.9% and 88.4±2.1%, respectively. The visibility is calculated by $V = (R_{max} - R_{min})/(R_{max} + R_{min})$, in which $R_{max}$ and $R_{min}$ are the maximum and minimum values of fitting. The accidental coincidence counts have been excluded. These fringes consist well with Eq. (2) except that it's not a clean separated state $|\Psi\rangle_{separated} = |1,1\rangle$ when $U=0$. From Fig. 2(b), a clean separated $|1,1\rangle$ state is obtained when $U_{offset}=2.3$V, namely the offset bias voltage. This may result from certain optical path difference, including asymmetry caused by etching the $SiO_2$ buffer layer and misalignment between the direction of inverted domain periodicity and the waveguides.

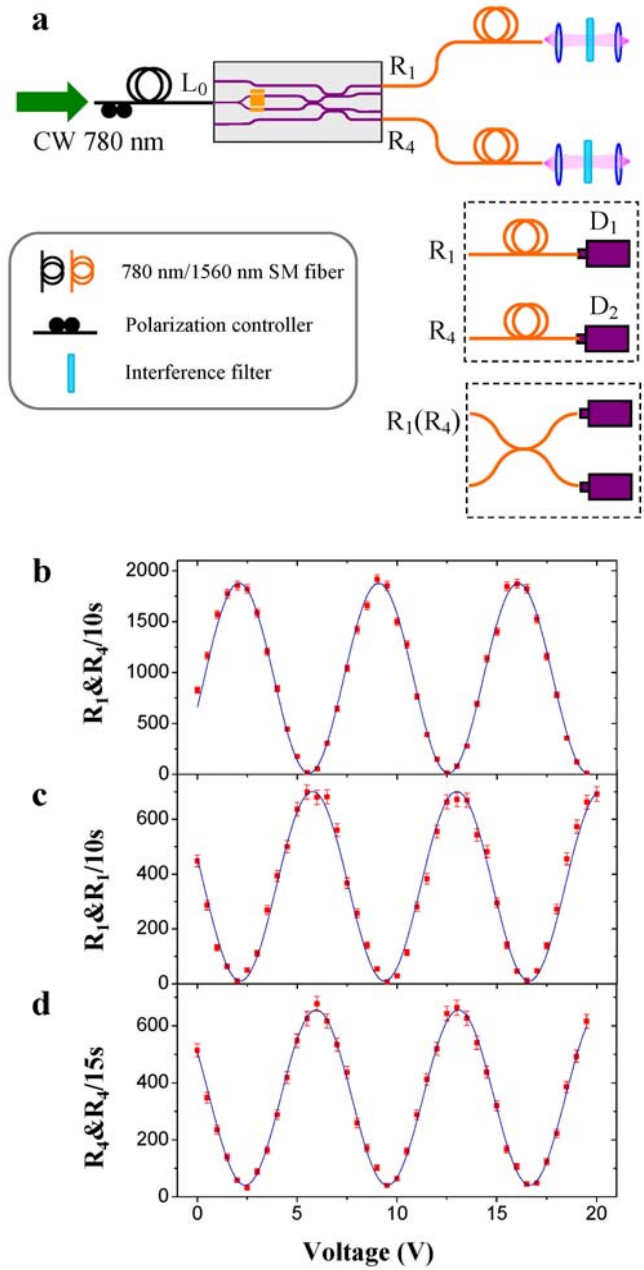

FIG. 2 On-chip two-photon interference. (a) Experimental setup. The fiber laser is polarization controlled and connected with the input fiber $L_0$. Before single-photon detectors, 14 nm bandwidth interference filters centered at 1560 nm are used to further eliminate the pump. The coincidence counting measurements are made between $R_1$&$R_4$, $R_1$&$R_1$ and $R_4$&$R_4$, respectively. The results are recorded in (b), (c) and (d). The solid lines are sinusoidal fitting and error bars show $\pm\sqrt{counts}$

The difference between two visibilities is mainly caused by imperfect coupling efficiency of $C_1$ (~54%) and different losses in waveguides 2 and 3. This coupling efficiency can be further optimized by applying a voltage to $C_1$. The small deviation between the positions of peaks in Fig. 2(b) and the dips in Figs. 2(c)-2(d) is attributed to long term direct-current (DC) drift associated with the buffer layer $(SiO_2)$[25,26]. Here the DC drift has been depressed a lot by etching a groove on the $SiO_2$ buffer layer as shown in the inset of Fig. 1. The waveguides are separated from the metal electrodes by the $SiO_2$ interlayer so as to reduce the propagation loss. The DC drift can be further minimized by replacing $SiO_2$ with a transparent conductive buffer layer such as indium tin oxide (ITO)[26].

According to the above measurement, we configure the chip to emit a clean $|\Psi\rangle_{separated}$ state by controlling the voltage at 2.3 V. To evaluate the quality of such state, we performe a Hong-Ou-Mandel (HOM) interference experiment by an external fiber beamsplitter. The experimental setup is sketched in Fig. 3(a). Photons from $R_1$ and $R_4$ are sent into into two fibers, one with a variable delay line in free space and the other with a polarization controller. Therefore, the indistinguishability of two photons in arrival time and polarization can be ensured when they interfere on the fiber beamsplitter. Figure 3(b) shows the coincidence counts that we recorded while varying the displacement of the free-space delay line. We can observe a HOM dip with a visibility of 92.9±0.9%. The nonideal visibility indicate that the $|\Psi\rangle_{separated}$ is

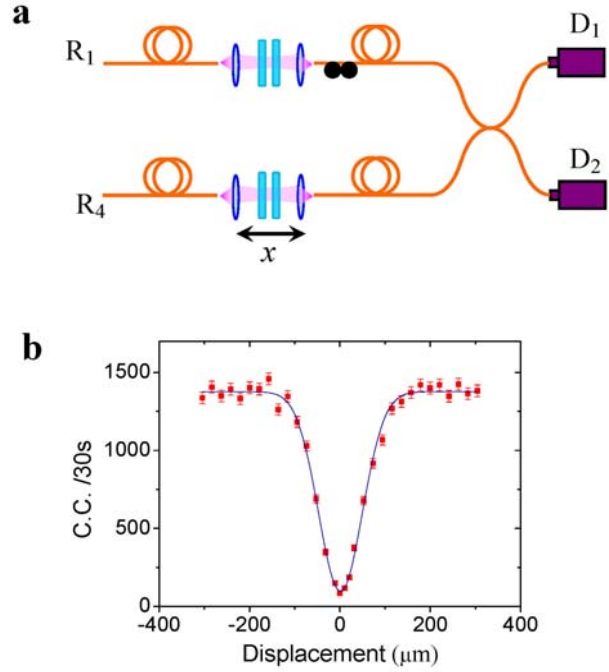

FIG. 3 Characterization of separated two-photon state by HOM interference. (a) Experimental setup. The photonic chip is controlled to emit separated photon-pair. Signal and idler photons are sent to two fibers, respectively. One is polarization controlled and the other is delayed by a variable free-space displacement before they interfere on a fiber beamsplitter. In each arm there are two interference filters of 14 nm bandwidth. (b) coincidence counts versus the displacement of the delay line. Error bars show $\pm\sqrt{counts}$

not absolutely clean, containing some bunched portion, as shown in Fig. 2(d).

To evaluate the conversion efficiency of our device, we configure the chip to emit separated state and measure the brightness of photon-pairs picked out by an interference filter with 14 nm bandwidth. The production rate of photon-pairs in the PPLN waveguides reached ~$6.2\times10^6$ Hz when the pump power in the input fiber was 39 μW. Taking into account the coupling efficiency of the pump from the fiber into the waveguide, which was about 80%, only ~31 μW was used. The production rate of photon-pairs per units bandwidth and pump power was ~$1.4\times10^7$ Hz nm$^{-1}$ mW$^{-1}$ for the 10 mm PPLN waveguide source.

The PPLN is the key section of this photonic chip, providing a flexible and feasible on-chip photon source. Here, by varying the period of PPLN, we actually designed 9 channels on a 3-inch LN wafer as shown in Fig. 4(a), intending to generate identical photon pairs whose wavelength covering the wavelength-division multiplexing (WDM) C-band (1530–1565 nm) and L-band (1565–1625 nm) in fiber communication. Each channel has the same waveguide circuits but different poling period, ranging from 14.36 μm to 16.28 μm in a step of 0.24 μm ( the corresponding SPDC process at room temperature will produce identical photon pairs around 1560 nm by ~12 nm separation). In this measurement, we verify the poling period and corresponding wavelength of identical photon pairs by second harmonic generation (SHG) in the assistant waveguide ($L_1$ or $L_4$). Figure 4(b) shows the wavelengths for efficient SHG in different channels at room temperature (25.5 °C), and they are also the wavelengths of identical photon pairs that can be generated by a pump laser with half the wavelength. In this work we choose the 5$^{th}$ channel whose period is 15.32 μm to produce the identical photon pair at 1560 nm. Figure 4(c) shows the whole structure for the channel. It contains three main photonic circuits, and each one has a structure similar to Fig. 1, including a Y branch splitter, electro-optical modulator, periodically poled section, 2×2 directional coupler ($C_1$) and filters ($C_2$ and $C_2'$). The difference lying among three main circuits is the interaction lengths of 2×2 directional coupler and filters. Different interaction lengths, including 650 μm/1300 μm, 750 μm/1500 μm and 850 μm/1700 μm, are designed to ensure a good splitting ratio for photon pairs which can be further optimized by other technologies like the coating or electro-optic modulation. The dashed rectangle region indicates the one we test in this experiment, with the interaction lengths of 650 μm for $C_1$ and 1300 μm for $C_2$ and $C_2'$. Besides, we also design some waveguide elements like single mode waveguide for 780 nm/1560 nm, directional coupler etc. (not shown) around the three main photonic circuits for testing and optimizing the waveguide fabrication technology. Figure 4(d) shows the photograph of our LN photonic chip with fixed fiber pigtails. From this chip, the two-color photon-pair is also accessible by increasing the working temperature due to the QPM condition of PPLN.

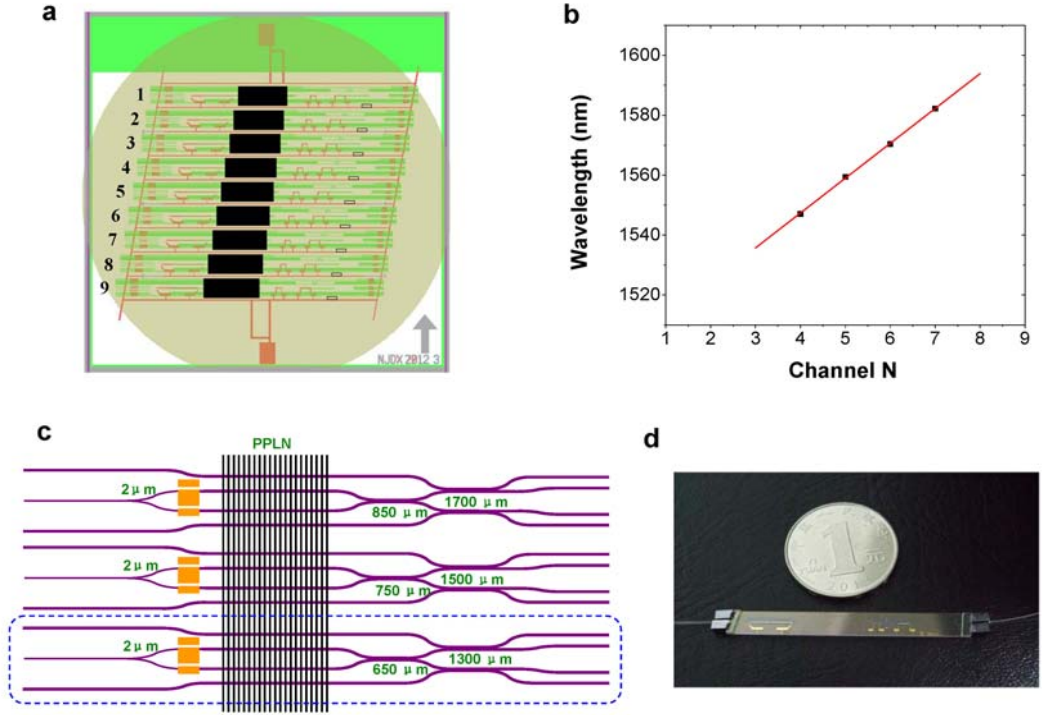

FIG. 4 LN photonic chip with multichannel design. (a) The layout of 9 channels with the same waveguide circuits but different inverted periods on a LN wafer. (b) Efficient SHG wavelengths in different channels at 25.5 °C. (c) The structure of the 5$^{th}$ channel with a poling period of 15.32 μm. It contains three main photonic circuits among which the difference only lie in the interaction lengths of 2×2 directional couplers (650 μm, 750 μm, 850 μm) and filters (1300 μm, 1500 μm, 1700 μm). The one we use is indicated by a dashed rectangle. Some extra waveguide elements are designed for convenient testing (not shown). The widths of single mode waveguides are 2 μm for the pump and 6 μm for photon pairs. (d) Photograph of one channel.

Recently a related work was published[11] which makes a great step toward fully integrated quantum optics by integrating entangled photons by four wave mixing (FWM) processes together with waveguide circuits on a silicon-on-insulator photonic chip. Silicon materials are considered to be competitive for quantum photonic devices for the mature fabrication technologies and accessible four wave mixing (FWM) photon sources[27-32]. But at the present stage, the nonlinearity and phase-modulator still need to be improved. As a contrast, although the LN circuits contains larger footprint and the fabrication technology is not compatible with the CMOS electronics, it contains efficient PPLN waveguide photon sources and fast electro-optic modulators, which makes it more competent for on-chip engineering of quantum light sources. By a careful design of quasi-phase-matching section, the polarization, frequency, spatial mode and path degrees of freedom can be engineered during the SPDC processes. When associated with phase-controlled circuits more types of quantum light sources can be exploited. This would definitely pave a way for complex quantum information processing. Besides, the LN photonic chip can act as an alternative platform for quantum walk especially when the nonlinear PPLN section is embedded[33,34], which may stimulate applications in quantum searching algorithms.